# Non-reciprocal robotic metamaterials

Martin Brandenbourger[1], Xander Locsin[1], Edan Lerner[1], Corentin Coulais[1]

[1]Institute of Physics, University of Amsterdam, Science Park 904, 1098 XH Amsterdam, The Netherlands.

**Abstract**: Non-reciprocal transmission of motion is potentially highly beneficial to a wide range of applications, ranging from wave guiding, to shock and vibration damping and energy harvesting. To date, large levels of non-reciprocity have been realized using broken spatial or temporal symmetries, yet only in the vicinity of resonances or using nonlinearities, thereby non-reciprocal transmission remains limited to narrow ranges of frequencies or input magnitudes and sensitive to attenuation. Here, we devise a novel type of robotic mechanical metamaterials wherein we use local control loops to break reciprocity at the level of the interactions between the unit cells. We show theoretically that first-of-their-kind asymmetric standing waves at all frequencies and unidirectionally amplified propagating waves emerge. We demonstrate experimentally and numerically that this property leads to tunable, giant, broadband and attenuation-free non-reciprocal performances, namely a level of 50dB non-reciprocal isolation over 3.5 decades in frequency, as well as one-way amplification of pulses.

**Main Text:**

Reciprocity is a fundamental property of linear, time-invariant physical systems, entailing that their response functions are symmetrical, namely that signals are transmitted symmetrically between any two points in space (*1*). In other words, if one sends an electromagnetic, acoustic or mechanical signal through a material in one direction, one can also send it in the opposite direction. Over the last few years, there has been an explosion of interest for breaking reciprocity in optics (*2-4*), microwaves (*5*), acoustics (*6*) quantum systems (*7, 8*) and mechanics (*9, 10*), thus creating new tools to engineer a novel generation of devices and materials that guide, damp or control energy and information in unprecedented ways. Non-reciprocity has been achieved using passive structures combining broken spatial symmetries and nonlinearities (*10, 11*) and using active time-modulated components that break time-reversal symmetry (*9, 12-15*). These strategies have led to large levels of non-reciprocal isolations, but with input magnitudes or input frequencies that are limited to narrow ranges, and are sensitive to attenuation. Here, inspired by recent developments in robotics (*16*) and active metamaterials (*12, 17, 18*) we introduce a novel robotic metamaterial (Fig. 1AB) that—through a combination of local sensing, computation, communication and actuation—uses distributed active-control to break reciprocity at the level of the interactions between the building blocks themselves. Such metamaterial features novel wave phenomena, namely asymmetric modes at all frequencies and unidirectional amplification and in turn realizes large, broadband, linear and self-amplified non-reciprocal transmission.

We first investigate theoretically the emergent properties in a mass-and-spring model with non-reciprocal springs (Fig. 1A). For reciprocal mechanical structures (*19-22*), the stiffness matrix—relating displacements to forces—is symmetrical by virtue of the Maxwell-Betti reciprocity theorem (*23, 24*). In particular, for a simple spring, left-to-right and right-to-left stiffnesses are equal $k_{L \to R} = k_{R \to L}$. Here, we consider a special mass-and-spring model, where the left-to-right and right-to-left stiffnesses differ $k_{L \to R} = k(1 + \varepsilon) \neq k_{R \to L} = k(1 - \varepsilon)$ (Fig. 1A). We obtain the following continuum equation (See Supplementary Text)

$$\frac{1}{c^2}\frac{d^2u}{dt^2} - \frac{d^2u}{dx^2} + \frac{2\varepsilon}{p}\frac{du}{dx} = 0, \tag{1}$$

where $c = p\sqrt{k/m}$ and where $p$ is the interparticle distance. In the case of reciprocal interactions ($\varepsilon=0$), Eq. (1) becomes the wave equation, which admits dispersion-free mechanical waves of group and phase velocity $c$. For non-reciprocal interactions $\varepsilon \neq 0$, the first order term in Eq. (1) breaks inversion symmetry $u \to -u$, $x \to -x$. This asymmetry has dramatic consequences on the nature of the mechanical waves, which can be readily seen from the solutions of this equation both in frequency domain and in real space. In frequency domain, solutions consist of a linear combination of the functions $\exp(i(\omega t - q_\pm x/p))$, where $q_\pm = i(\varepsilon \pm \sqrt{\varepsilon^2 - \omega^2 p^2/c^2})$. For small frequencies $|\omega| < c|\varepsilon|/p$, these solutions are exponentially localized standing waves, while for large frequencies $|\omega| > c|\varepsilon|/p$, they are localized oscillatory standing waves with an exponential envelope (Fig. 1B). Crucially, for $\varepsilon > 0$ ($\varepsilon < 0$) the imaginary part of $q_\pm$ is always positive (negative) so these solutions are always localized on the right (left) edge. In real space, we obtain the Green's function of Eq. (1) (see Supplementary Text), which is an asymmetric step function propagating at a velocity $c$ with a wave front magnitude given by $c\exp[\varepsilon c t/p]/2$ ($c\exp[-\varepsilon c t/p]/2$) for $x > 0$ ($x < 0$). For any value of $\varepsilon > 0$ ($\varepsilon < 0$), the initial pulse is amplified for forward (backward) propagation and attenuated for backward (forward) propagation (Fig. 1C). Therefore, the $\frac{2\varepsilon}{p}\frac{du}{dx}$ term that breaks inversion-symmetry in Eq. (1) leads to waves with two unprecedented features, namely spatial asymmetry at all frequencies and unidirectional amplification.

In order to create a system with such non-reciprocal local interactions, it is important to realize that breaking translational invariance by adding local external forces is necessary. To do so, we built a metamaterial made of ten "robotic" building blocks (Fig. 2A). Each robotic unit cell consists of a mechanical rotor with a rotational moment of inertia $J$, of a local control system and is mechanical coupled to its neighbors via pre-stretched elastic beams resulting in a torsional stiffness $C$ (Fig. 2BC). The control system measures the rotor's angular position $\theta_L$ and collects that of its right neighbor $\theta_R$, and applies an additional torque $\tau_L^M = C f(\alpha) (\theta_L - \theta_R)$ on the rotor. $\alpha$ is a non-dimensional feedback parameter and $f(\alpha)$ is a gain function (See Supplementary Text). We calibrate the torque vs. angle response between two unit cells and find, as expected, that $C_{L \to R} = C$ differs from $C_{R \to L} = C(1 - f(\alpha))$ (Fig. 2D). While such tunable non-reciprocal response is not surprising—ultimately it is achieved at the level of each unit cell's microcontroller—the novelty of our approach lies in coupling many such robotic non-reciprocal unit cells together.

To test the predictions of the mass-and-spring model, we now investigate experimentally and numerically the stationary response of our ten-unit cells robotic metamaterial to harmonic excitations on the left and on the right edges over a wide range of input frequencies (see Supplementary Text for details). In the reciprocal case α=0 (Fig. 3A), we observe that the angular displacement profiles either decay exponentially or oscillate. The responses to left and right excitations are simply related by mirror symmetry, which demonstrates that the metamaterial response is inherently symmetrical. In contrast, in the non-reciprocal case α=0.43 (Fig. 3B), we observe a strong asymmetry in the angular displacement profiles. When excited from the right, the response is more localized close to the excitation point and when excited from the left, the response is more extended towards the right and even increases for large frequencies. This asymmetry is further quantified by the spatial decays of the profiles, which are opposite in the reciprocal case α=0 and differ in the non-reciprocal case α=0.43 (Fig. 3C), regardless of the driving frequency. We therefore observe asymmetric modes at all frequencies, as predicted by the solutions of Eq. (1).

Does such strong asymmetry lead to non-reciprocal transmission? To address this question, we calculate the transmissions $T_{L \to R}(f) = 20 \log_{10} \left| \bar{\theta}_R^{out} / \bar{\theta}_L^{in} \right|$ and $T_{R \to L}(f) = 20 \log_{10} \left| \bar{\theta}_L^{out} / \bar{\theta}_R^{in} \right|$ from the angular displacements profiles obtained above. In the reciprocal case ($\alpha$=0), we observe a symmetrical transmission, typical of a resonant low-pass filter, with a saturated plateau of amplitude –37 +/- 1 dB at small frequencies and a broad peak corresponding to the system resonances beyond which the transmission signal starkly decreases at 360dB/decade (Fig. 2E). For the non-reciprocal case ($\alpha$=0.43), the left-to-right and right-to-left transmissions differ vastly—by more than 50dB—over a wide range of frequencies, from 0.001 to 5Hz (Fig. 2F). Importantly, such non-reciprocal isolation increases with the system size (Fig. 2F-inset). Therefore, the existence of the asymmetric modes at all frequencies leads to an extremely large level of non-reciprocal transmission over a very large range of frequencies, a performance that is unprecedented amongst wave-based physical systems.

Since the non-reciprocal transmission is broadband, our robotic metamaterial is in principle an excellent non-reciprocal device for pulses, which have a broadband spectral signature. To demonstrate this, we excite our metamaterial with half-sine shaped pulses of amplitude 0.04 rad and duration 100 ms, either from the left or right edge. In the reciprocal case $\alpha$=0, the response is strictly the same regardless of the excitation point (Fig. 3A-C). The pulses propagate across the metamaterial, reflect on the opposite edge, propagate back, reflect on the initial edge, and so on. After a short transient and before the first reflection, the pulse amplitude decreases. By contrast, in the non-reciprocal case $\alpha$=0.62 (Fig. 3D-F), the pulse attenuates strongly when excited from the right and is amplified when excited from the left. Such unidirectional amplification is tunable (see Fig. 4F-inset) and is in qualitative agreement with the behavior of the Green's function of Eq. (1) discussed above.

To conclude, we have devised a novel class of robotic mechanical metamaterials, that are embedded with non-reciprocal interactions through local control-loops. As a result, they feature a completely novel type of wave phenomena, which show spatial asymmetry of standing waves at all frequencies leading to unprecedented broadband giant level of non-reciprocity and which exhibit unidirectional amplification of pulses. These non-reciprocal robotic metamaterials offer new vistas for applications where unidirectional transmission of energy is useful, e.g. for communication and sensing (*3, 6, 13, 18, 25*), shock and vibration damping and energy harvesting (*17, 26*). Our study opens up a plethora of future research directions, e.g. the investigation of instabilities (*27*) and exceptional points (*28*), miniaturization, and generalization in higher dimensions, and beyond mechanics for acoustics, flexural waves photonics and quantum systems.

**Acknowledgments:** We thank D. Giesen, G. Hardeman, U. van Hes, T. Walstra and T. Weijers for their skillful technical assistance. We are grateful to A. Alù and J. van Wezel for insightful discussions. C.C. acknowledges funding from the Netherlands Organization for Scientific research (NWO) VENI grant No. NWO-680-47-445.

Figures:

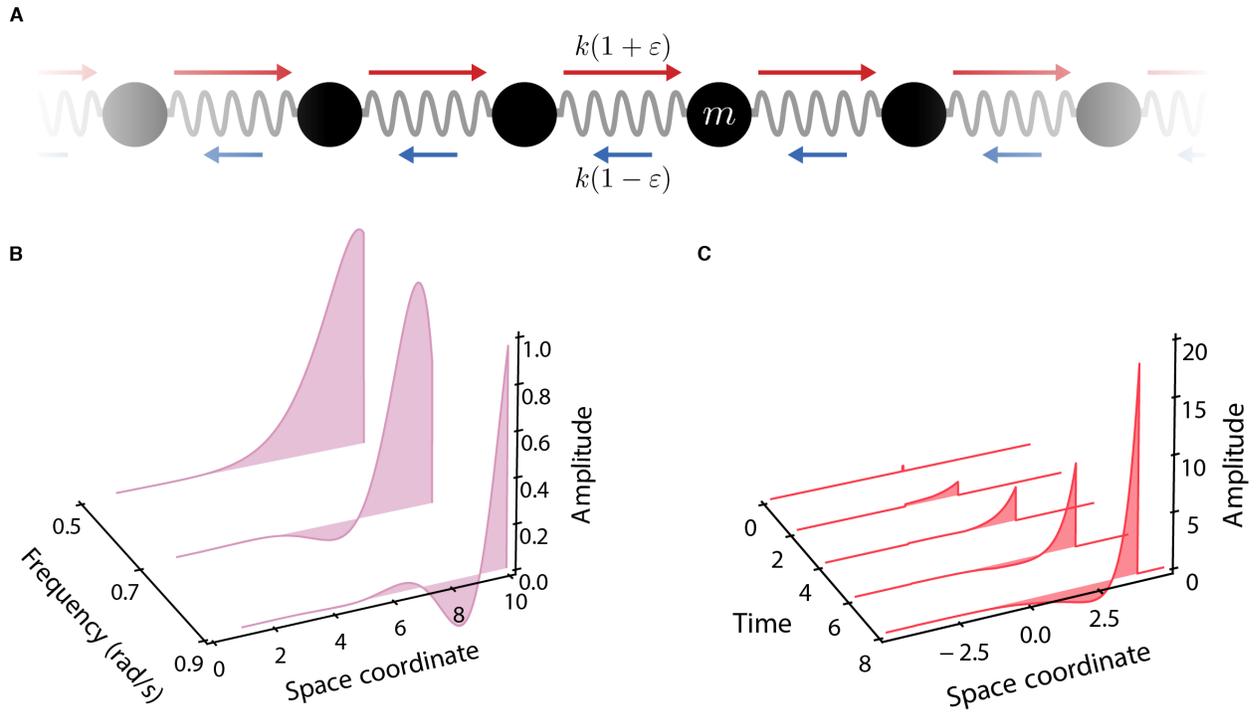

**Fig. 1**. **Asymmetric and unidirectionally amplified waves in a non-reciprocal mass-and-spring model**. (**A**) Schematic representation of the non-reciprocal mass-and-spring model. (**B**) Magnitude of the solutions of Eq. (1) in the frequency domain $\exp(i(\omega t - q_{\pm}x/p))$ vs. space coordinate $x$, for three different frequencies. (**C**) Green's function of Eq. (1) vs. time and spatial coordinates. In (**B**) and (**C**) $\varepsilon = 0.9$ and c = 0.5.

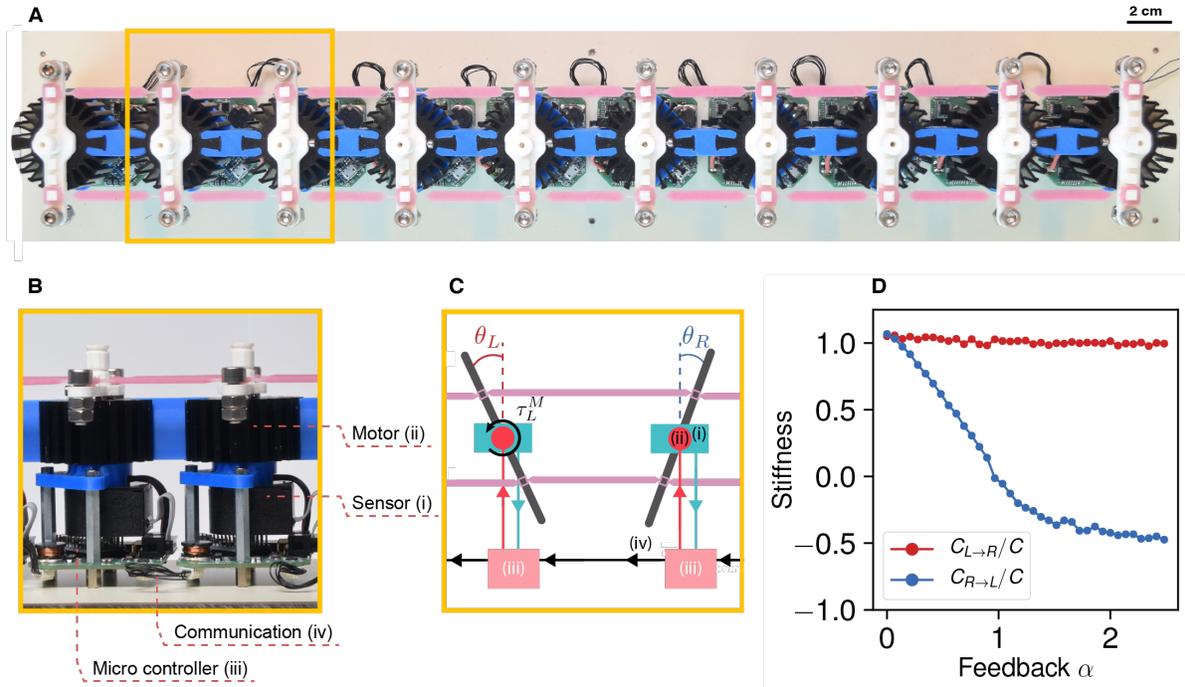

**Fig. 2**. **Robotic metamaterial with non-reciprocal interactions.** (**A**) Robotic metamaterial made of 10 unit cells mechanically connected by soft elastic beams. (**BC**) Close-up (**B**) and sketch (**C**) on two unit cells. Each unit cell is a minimal robot with a unique rotational degree of freedom that comprises an angular sensor (i), a coreless DC motor (ii), and a microcontroller (iii). Each unit cell communicates with its right neighbor via electric wires (iv). These components allow to program a control loop characterized by the feedback parameter α (see main text for definition). (**D**) Rescaled torsional stiffnesses $C_{L \to R}/C$ (red) and $C_{R \to L}/C$ (blue) as a function of the feedback parameter $\alpha$.

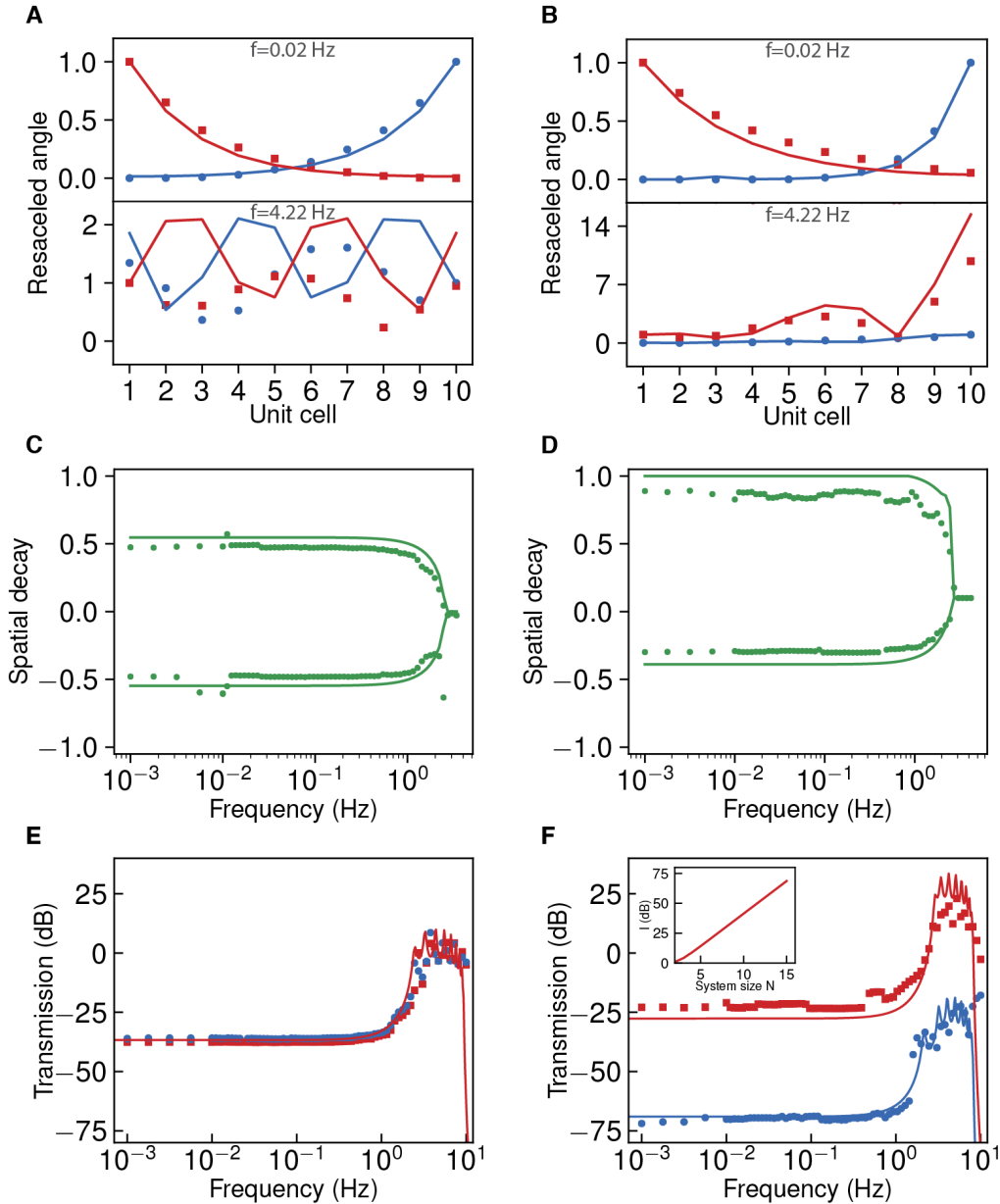

**Fig. 3**. **Asymmetric standing waves and broadband unidirectional transmission.** Stationary response under sinusoidal excitations at the left (red squares) or right (blue dots) edge of the metamaterial. (**AB**) Amplitudes of oscillation rescaled by the amplitude of actuation as a function of the unit cell position for sinusoidal excitations with α=0 (**A**) and α=0.43 (**B**) at frequencies 0.02 Hz (top) and 4.22 Hz (bottom). See also Supplementary Movie 1. (**CD**) Spatial

decay rates of the amplitudes of oscillation vs. frequency with α=0 (**C**) and α=0.433 (**D**). (**EF**) Left-to-right ($T_{L \to R}$) and right-to-left ($T_{R \to L}$) transmissions as a function of the excitation frequency (f) for a feedback parameter α=0 (**E**) and α=0.433 (**F**). (**F**-inset) Isolation at low frequency $I = T_{L \to R} - T_{R \to L}$ as a function of the number of unit cells N, for α=0.43. For each graph, markers depict the experiments and solid lines the numerical model that has been independently calibrated without any fit parameters (see Supplementary Text). The agreement is very good until 3Hz, above which the numerical model is too simplistic to accurately capture internal vibrations of the rubber bands. Besides, note that beyond 10Hz our experimental system cannot drive and measure.

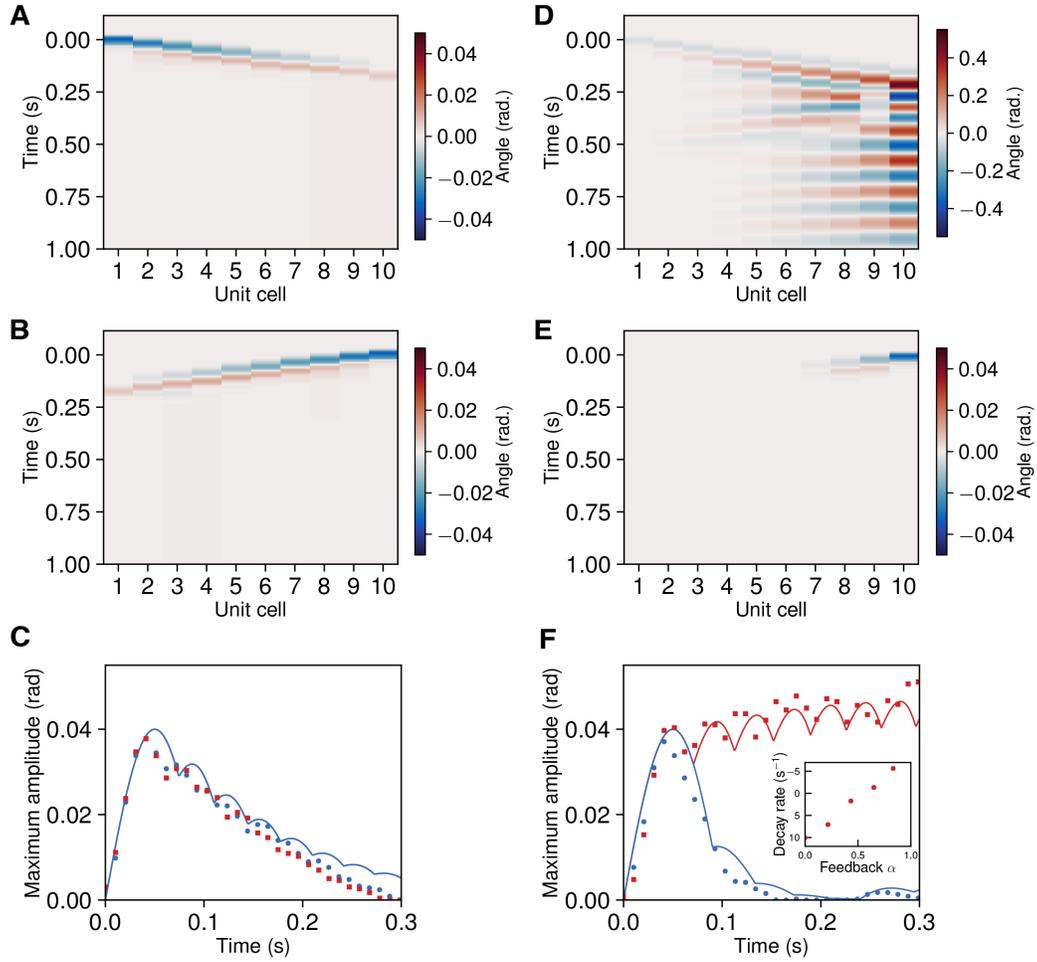

**Fig. 4**. **Unidirectional pulse amplification.** (**A-E**) Contour plots of the angular displacement vs. space coordinate and time for a feedback parameter α=0 (**AB**) and α=0.62 (**DE**), upon pulse excitation on the left (**AD**) and right (**BE**) edge of the metamaterial. (**CF**) Instantaneous maximum magnitude of the propagation pulse vs. time for a feedback parameter α=0 (**C**) and α=0.62 (**F**). The points correspond to the experimental data, the thin solid lines to the numerical model without any fit parameters. (**F**-inset) Decay rate δ derived from exponential fits of the form exp(−δ t) on the maximum amplitude of pulses propagating from left to right for different values of feedback parameter α. See also Supplementary Movie 2.